# Deep Autoencoder-based Fuzzy C-Means for Topic Detection

Hendri Murfi
Department of Mathematics
Universitas Indonesia, Depok 16424, Indonesia
Email: hendri@ui.ac.id

Natasha Rosaline
Department of Mathematics
Universitas Indonesia, Depok 16424, Indonesia
Email: natasha.rosaline@sci.ui.ac.id

Nora Hariadi
Department of Mathematics
Universitas Indonesia, Depok 16424, Indonesia
Email: nora.hariadi@sci.ui.ac.id

**Abstract**. Topic detection is a process for determining topics from a collection of textual data. One of the topic detection methods is a clustering-based method, which assumes that the centroids are topics. The *clustering* method has the advantage that it can process data with negative representations. Therefore, the clustering method allows a combination with a broader representation learning method. In this paper, we adopt deep learning for topic detection by using a deep autoencoder and fuzzy c-means called deep autoencoder-based fuzzy c-means (DFCM). The encoder of the autoencoder performs a lower-dimensional representation learning. Fuzzy c-means groups the lower-dimensional representation to identify the centroids. The autoencoder's decoder transforms back the centroids into the original representation to be interpreted as the topics. Our simulation shows that DFCM improves the coherence score of eigenspace-based fuzzy c-means (EFCM) and is comparable to the leading standard methods, i.e., nonnegative matrix factorization (NMF) or latent Dirichlet allocation (LDA).

## 1. Introduction

Topic detection is a process used to analyze words in a collection of textual data to determine the topics in the collection, how they relate to one another, and how they change over time. The topics usually are represented by a set of words. The coherence of the words usually measures the topic's interpretability. The standard topic detection methods are *nonnegative matrix factorization* (NMF) (Lee & Seung, 1999), *clustering* (Allan, 2002), and *latent Dirichlet allocation* (LDA) (Blei et al., 2003).

In the *clustering* method, the cluster centers or centroids are interpreted as a topic. In other words, the clustering method will group the textual data based on their topic similarity. Unlike the other two methods, the *clustering* method can process data with negative representations. Therefore, the clustering method allows a combination with broader representation learning or dimension reduction methods. Nur' aini et al. combines *k-means* and *latent semantic analysis* (LSA) for topic detection (Nur'aini, Najahaty, Hidayati, Murfi, & Nurrohmah, 2015). Firstly, the textual data are transformed into a lower-dimensional Eigenspace using the singular value decomposition (SVD). Next, *k-means* is performed on the Eigenspace to extract the topics that are then transformed back to the nonnegative subspace of the original space.

The *k-means* method splits the textual data into *k* clusters in which each textual data belongs to the nearest centroid. It means that the *k-means* method assumes that each textual data contains only one topic. This assumption is relatively weak and also different from the standard NMF and LDA, considering that the textual data may have many topics. Therefore, soft clustering is examined to be an alternative clustering method for topic detection. *Fuzzy c-means* (FCM) is one of the famous soft clustering methods (Bezdek, Ehrlich, & Full, 1984). Using FCM, the textual data may belong to more than one cluster and may have more than one topic. The combination of FCM and LSA called *Eigenspace-based fuzzy c-means* (EFCM) is proposed for topic detection (Muliawati & Murfi, 2017). In general, some simulations show that EFCM gives the coherence scores between the ones of LDA and NMF (Murfi, 2018, 2019; Praditya Nugraha, Rifky Yusdiansyah, & Murfi, 2019).

Currently, deep learning is the primary machine learning method for unstructured data such as images and text (Goodfellow, Bengio, & Courville, 2016; Zhang, Lipton, Li, & Smola, 2020). Deep learning has been extensively studied to extract a good representation of data by neural networks (Bengio, Courville, & Vincent, 2013). In this paper, we adopt deep learning to improve the performance of EFCM for topic prediction problems by using deep autoencoder (DAE) for the representation learning process. We call this topic detection method as deep autoencoder-based fuzzy c-means (DFCM). First, the encoder of DAE performs a lower-dimensional representation learning. Next, FCM groups the lower-dimensional representation to identify the centroids. Finally, the decoder of DAE transforms back the centroids into the original representation to provide the topics. Our simulation shows that DFCM improves the coherence score of EFCM and is comparable to the leading standard methods, i.e., NMF or LDA.

This paper's outline is as follows: In Section 2 and Section 3, we describe the related works and the methods, i.e., FCM, DAE, and DFCM. Section 4 describes the results and the discussion of our simulations. Finally, a general conclusion about the results is presented in Section 5.

## 2. Related Works

Topic detection methods are algorithms for discovering the topics or the themes from an unstructured collection of documents. Some recent publications show the gwowing use of topic detection for researchers in Library and Information Science to find the theme that they are interested in and then examine the documents related to that theme (Battsengel, Geetha, & Jeon, 2020; Lamba & Madhusudhan, 2019; Parlina, Ramli, & Murfi, 2020).

The standard topic detection methods are *nonnegative matrix factorization* (Cichocki & Phan, 2009; Févotte & Idier, 2011; Lee & Seung, 1999), *clustering* (Allan, 2002; Petkos, Papadopoulos, & Kompatsiaris, 2014) and *latent Dirichlet allocation* (Blei, 2012; Blei et al., 2003; Hoffman, Blei, & Bach, 2010; Hoffman, Blei, Wang, & Paisley, 2013). Unlike the other two methods, clustering is a general method of grouping data. Furthermore, this method can also process positive and negative data representation. Thus, the clustering method is more flexible to be combined with representation learning or dimension reduction.

Fuzzy clustering is one of the most widely used clustering methods because it has soft and flexible in grouping data to the cluster (Ruspini, Bezdek, & Keller, 2019). Bezdek developed FCM by extending the fuzzifier value m to m > 1 (Bezdek et al., 1984). This extension makes FCM a generalization from k-means, which is hard clustering. FCM is more suitable for the topic detection method because it allows adaptation to a document's condition with one or more topics, namely by finding the optimal fuzzifier value m.

In the era of big data, the existence of high-dimensional data is a big challenge for FCM (Winkler, Klawonn, & Kruse, 2011). By finding a new representation of the original data, two approaches are already used to reduce the difficulty of FCM in high-dimensional data. The first approach uses kernel methods to implicitly get more expressive features by formulating the data into the feature space constructed by some kernel functions (Huang, Chuang, & Chen, 2012; Shang, Zhang, Li, Jiao, & Stolkin, 2019). The second approach is an explicit transformation of the original data. In addition to the specified nonlinear data transformations (Zhu, Pedrycz, & Li, 2017), random projection (Rathore, Bezdek, Erfani, Rajasegarar, & Palaniswami, 2018) is commonly used to obtain low-dimensional data. The second approach is more suitable because the excellent design of kernel space for clustering is complicated. The ability of kernel methods to handle large-scale data is also a concern. In several empirical studies, the combination of FCM with the data transformation approach (Murfi, 2018; P. Nugraha, Rifky Yusdiansyah, &

Murfi, 2019) provides better performance than the random projection approach (Yusdiansyah, Murfi, & Wibowo, 2019) for topic detection problems.

Currently, deep learning is the primary machine learning method for unstructured data such as images and text (Goodfellow et al., 2016; Zhang et al., 2020). Deep learning has been extensively studied to extract a good representation of data by neural networks (Bengio et al., 2013). The combination of the deep neural network and an unsupervised clustering method also becomes an active research field (Song, Huang, Liu, Wang, & Wang, 2014). In general, there are some approaches to incorporate deep learning. In general, there are several approaches to combining deep learning and clustering. The first approach is to combine representation learning and clustering in two steps, namely using a deep autoencoder for representation learning and then a clustering method for the next stage (Song et al., 2014; Song, Liu, Huang, Wang, & Tan, 2013). The second approach is to combine a deep autoencoder and a clustering method simultaneously (Guo, Gao, Liu, & Yin, 2017; Xie, Girshick, & Farhadi, 2016). The next approach is to combine clustering with a pretrained encoder, such as Bidirectional Encoder Representations from Transformers (BERT) proposed by Google (Guan et al., 5555). However, most of the clustering methods used in these approaches are hard clustering. Few studies work on the improvement of feature quality by deep learning for the fuzzy clustering. This study is to find a good deep representation for the fuzzy clustering, i.e. fuzzy c-means. In this research, we use the first approach and the second approach in combining representation learning and clustering. In this approach, representation learning and clustering are carried out separately, not simultaneously as in the second approach. This approach still requires the decoder part to transform the data back into original representation. In addition, our approach does not use a pretrained model because it is still difficult to determine the most important words to represent the resulting topics.

## 3. Methods

Let $A$ be a word by document matrix and $c$ be the number of topics. Given $A$ and $c$, the topic detection problem is how to recover $c$ topics from $A$. In the clustering-based topic detection method, the clustering centers or centroids are interpreted as topics. In this section, we describe *deep autoencoder-based fuzzy c-means* (DFCM) for topic detection. First, we review the core methods, i.e., *fuzzy c-means* (FCM) and *deep autoencoder* (DAE).

### 3.1. Fuzzy C-Means

Given a dataset in the form of a word by document matrix $A = [\boldsymbol{a}_1\ \boldsymbol{a}_2\ ...\ \boldsymbol{a}_n]$ and the number of centroids $c$, the goal of *fuzzy c-means* (FCM) can be formulated as the following constrained optimization:

$$\min_{m_{ik},\boldsymbol{q}_i} J = \sum_{i=1}^{c}\sum_{k=1}^{n} m_{ik}^{f} \|\mathbf{a}_k - \mathbf{q}_i\|^2 \qquad (1)$$

$$s.t. \quad \sum_{i=1}^{c} m_{ik} = 1, \forall k$$
$$0 < \sum_{k=1}^{n} m_{ik} < n, \forall i$$
$$m_{ik} \in [0,1], \forall i,k$$

where $\boldsymbol{q}_i$ are centroids, $m_{ik}$ is the membership of data point $\boldsymbol{a}_k$ in cluster $i$, $f > 1$ is the fuzzification constant, and $\|\ .\ \|$ is any norm. The first constraint ensures that every data point has total membership in all clusters where each membership is in [0,1]. The second constraint guarantees that all clusters are nonempty(Bezdek et al., 1984).

The problem of the constrained optimization in Equation 1 is to find $m_{ik}$ and $\boldsymbol{q}_i$ that minimize the objective function $J$. The standard method to solve the constrained optimization is alternating optimization. First, we choose some initial values for the $\boldsymbol{q}_i$. Then we minimize $J$ concerning the $m_{ik}$, keeping the $\boldsymbol{q}_i$ fixed giving:

$$m_{ik} = \left[\sum_{j=1}^{c}\left(\frac{\|\mathbf{a}_k-\mathbf{q}_i\|}{\|\mathbf{a}_k-\mathbf{q}_j\|}\right)^{2/f-1}\right]^{-1}, \forall i,k \qquad (2)$$

Next, we minimize $J$ on the $\boldsymbol{q}_i$, keeping the $m_{ik}$ fixed giving:

$$\mathbf{q}_i = \frac{\sum_{k=1}^{n}((m_{ik})^f \mathbf{a}_k)}{\sum_{k=1}^{n}(m_{ik})^f}, \forall i \qquad (3)$$

This two-step optimization are iterated until a stopping criterion is fulfilled, e.g., the maximum number of iteration, insignificant changes in the objective function $J$, the membership $m_{ik}$, or the centroids $\boldsymbol{q}_i$ (Bezdek & Hathaway, 2003). The FCM algorithm is described in more detail in Algorithm 1.

According to Equation 2, the memberships $m_{ik}$ tend to *0* or *1* when the fuzzification constant $f$ approaches to 1. The bigger the fuzzification constant makes, the fuzzier the memberships $m_{ik}$. Therefore, the setting of the fuzzification constant is quite intuitive. The small fuzzification constant means that each textual data may contain a small number of topics. On the other hand, the bigger the fuzzification constant implies that each textual data may have more topics.

**Algorithm 1.** FCM
Input   : $A$, $c$, f, max iteration ($T$), threshold ($\varepsilon$)
Output  : $\boldsymbol{q}_i$
1. set $t = 0$
2. initialize $\boldsymbol{q}_i$
3. update $t = t + 1$
4. calculate $m_{ik} = \left[\sum_{j=1}^{c}\left(\frac{\|\mathbf{a}_k-\mathbf{q}_i\|_2}{\|\mathbf{a}_k-\mathbf{q}_j\|_2}\right)^{2/f-1}\right]^{-1}, \forall i,k$
5. calculate $\boldsymbol{q}_i = \frac{\sum_{k=1}^{n}((m_{ik})^f \mathbf{a}_k)}{\sum_{k=1}^{n}(m_{ik})^f}, \forall i$
6. if a stopping, i.e., $t > T$ or $\|M^t - M^{t-1}\|_F < \varepsilon$, is fulfilled then stop, else go back to step 3

### 3.2. Deep Autoencoder

*Deep autoencoder* (DAE) is a deep neural network for unsupervised learning problems. This unsupervised problem is solved using a supervised learning approach, where target labels are constructed from input features. This deep autoencoder architecture has the same output layer as the input layer, and the standard supervised learning can be applied.

The architecture of DAE for representation learning can be explained into three parts, i.e., encoder, code, and decode (Figure 1). The encoder part consists of fully connected layers used to transform data input to the code part, a new data representation. The decoder part is used to transform the new data representation back to the original representation. The decoder part consists of fully connected layers having a symmetric structure with the encoder part. The reason is if an encoder requires a certain complexity (number of layers (depth), the number of neurons in each layer (units)) to represent data to new representation, then a decoder with the same complexity is needed to transform the new data representation back to the original data representation. For dimension reduction problem, the number of neurons in the code part is set less than the number of neurons in the input layer (Hinton & Salakhutdinov, 2006).

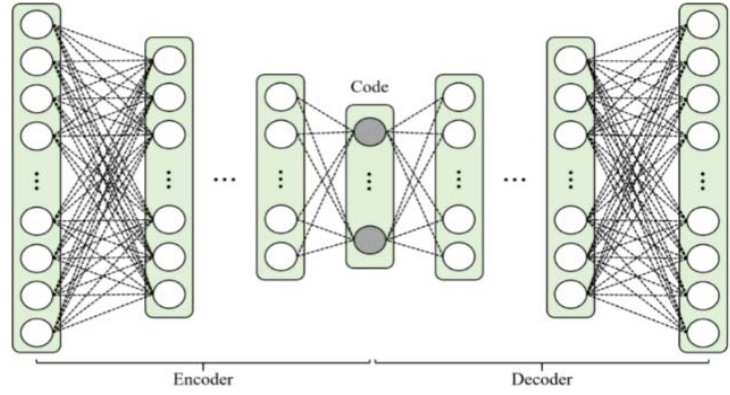

Figure 1. Deep Autoencoders

DAE can be built layer by layer using greedy layer-wise pretraining, where each layer is built by a denoising autoencoder (Vincent et al., 2010). The denoising autoencoder is an autoencoder that reconstructs the input from a corrupted version to force the hidden layer to discover a more stable and robust representation.

Given textual dataset $X = \{\mathbf{x}_1, \mathbf{x}_2, \dots, \mathbf{x}_N\}$ with $\mathbf{x}_i \in \mathcal{R}^{D_x}$, $\forall i = 1,2, \dots, N$, the denoising autoencoder consists of two layers as follows:

$$\widetilde{\boldsymbol{x}}_i \sim dropout_1(\boldsymbol{x}_i) \tag{4}$$

$$\boldsymbol{h}_i = g_1(\widetilde{\boldsymbol{x}}_i, \boldsymbol{w_1}) \tag{5}$$

$$\widetilde{\boldsymbol{h}}_i \sim dropout_2(\boldsymbol{h}_i) \tag{6}$$

$$\boldsymbol{y}_i = g_2(\widetilde{\boldsymbol{h}}_i, \boldsymbol{w_2}) \tag{7}$$

where $dropout_1()$ and $dropout_2()$ are methods to ignore some number of neuron outputs during training randomly, $g_1()$ and $g_2()$ are activation function, $\boldsymbol{w}_1$ and $\boldsymbol{w}_1$ are weights. The fitting is performed to minimize the loss $\mathcal{L}(\boldsymbol{x}, \boldsymbol{w})$, i.e., errors between $\boldsymbol{x}_i$ and $\boldsymbol{y}_i$. Next, $\boldsymbol{h}_i$ becomes a new representation for the input data of the next layer. After training on each denoising autoencoder, the denoising autoencoder's weights become the corresponding weights of the autoencoder. Furthermore, the autoencoder is retrained to minimize a reconstruction loss for all layers. This DAE algorithm is described in more detail in Algorithm 2.

3.3. Deep Autoencoder-based Fuzzy C-Means

*Deep autoencoder-based Fuzzy C-Means* (DFCM) is a proposed topic detection method that combines DAE for representation learning and FCM for fuzzy clustering. FCM works well for low dimensional textual data and generates only one topic for high dimensional textual data. We can set the fuzzification constant of FCM with a small value to push FCM to produce more than one topic. However, this small fuzzification constant assumes that each textual data contains few topics and only one topic when the fuzzification constant approaches to one. Therefore, we use DAE to transform the data to lower-dimensional representation and keep the fuzzification constant adaptable for textual data with multi-topics. Figure 2 provides a general process of DFCM.

**Algorithm 2.** DAE
Input  : $X$, the size of code $p$
Output: $encoder(w)$, $decoder(w)$
1. Initialize $autoencoder(p)$
2. $h_n^{(0)} = x_n$
3. Let $m$ be the number of layers of the autoencoder
4. FOR $i = 1$ TO $m$
5.     Fitting the denoising autoencoder for the $i$-th layer:
$$deAutoencoder(w^{(i)}), w^{(i)} = \min_{w} \mathcal{L}\left(h_n^{(i-1)}, w\right), \forall n$$
6.     $h_n^{(i)} = deEncoder\left(h_n^{(i-1)}\right), \forall n$
7. Initialize weight of autoencoder with the corresponding weight of the denoising autoencoder: $autoencoder(w^{(i)}), \forall i$
8. Fitting the autoencoder: $autoencoder(w), w = \min_{w} \mathcal{L}(x_n, w), \forall n$

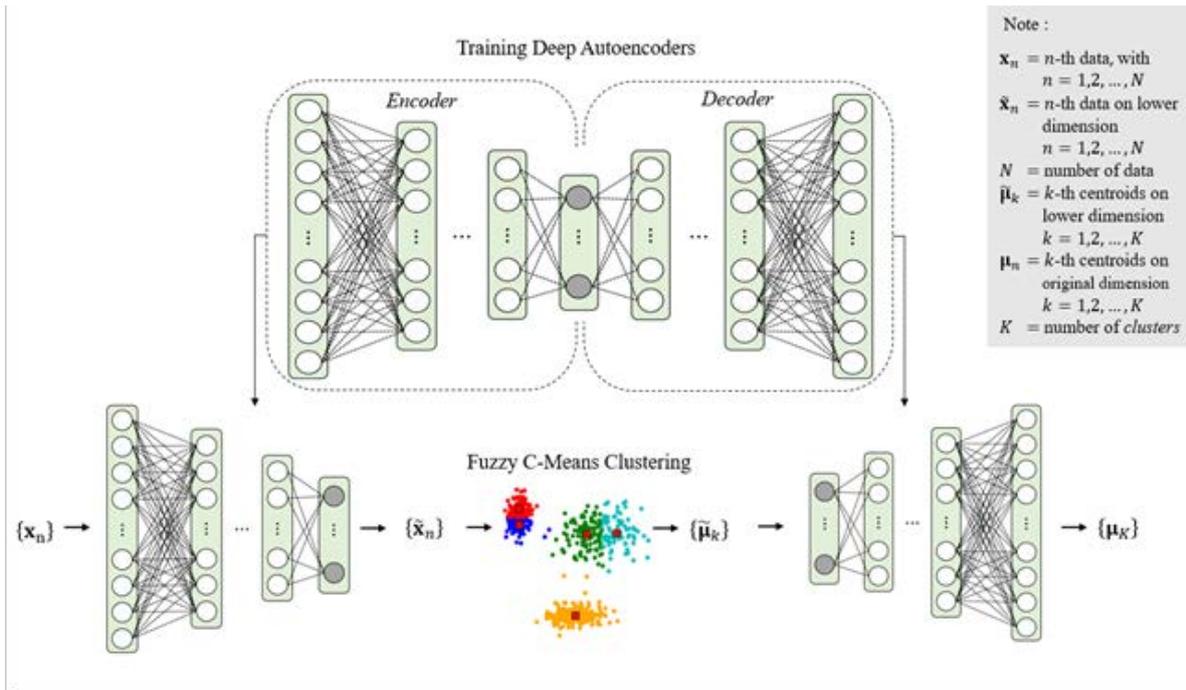

Figure 2. Deep Autoencoders-based Fuzzy C-Means

Given a textual dataset $X = \{\mathbf{x}_1, \mathbf{x}_2, \ldots, \mathbf{x}_N\}$ with $\mathbf{x}_i \in \mathcal{R}^{D_x}$, $\forall i = 1,2,\ldots,N$, the dimension of new data representation $p$, and the number of topic $c$. First, the textual data are transformed into a lower-dimensional representation using an encoder. We denote this transformation as follows:

$$\tilde{X} = encoder(X, p) \tag{8}$$

where $\tilde{\mathbf{x}}_i \in \mathcal{R}^{D_{\tilde{x}}}$, $\forall i = 1,2,\ldots,N$. Next, we perform FCM on the dataset $\tilde{X}$ with the lower-dimensional representation. In this step, centroids $\tilde{\boldsymbol{\mu}}_i \in \mathcal{R}^{D_{\tilde{x}}}$, $\forall i = 1,2,\ldots,c$ are extracted from all $c$-given clusters as follows:

$$\tilde{\boldsymbol{\mu}}_i = FCM(\tilde{X}, c, f, T, \varepsilon) \tag{9}$$

These centroids $\tilde{\boldsymbol{\mu}}_i$ are interpreted as the topics in the lower-dimensional representation. However, the topics have no meaning and will be meaningful if they are transformed back to the original representation. Therefore, it is necessary to transform the extracted topics back to the original representation as follows:

$$\boldsymbol{\mu}_i = \max(0, decoder(\tilde{\boldsymbol{\mu}}_i)) \tag{10}$$

where $\boldsymbol{\mu}_i \in \mathcal{R}^{D_x}$, $\forall i = 1,2,\ldots,c$, and max() is a function that gives a maximum between 0 and each element of $decoder(\tilde{\boldsymbol{\mu}}_i)$. This DFCM algorithm is described in more detail in Algorithm 3.

---

**Algorithm 3.** DFCM

Input : $X$, the size of code $p$, the number of topics $c$, max number of iterations T, threshold $\varepsilon$

Output: $\boldsymbol{\mu}_i$
1. Build autoencoder: $encoder, decoder = DAE(X, p)$
2. Transform X : $\tilde{X} = encoder(X)$
3. Perform FCM : $\tilde{\boldsymbol{\mu}}_i = FCM(\tilde{X}, c, T, \varepsilon), i = 1,2,\ldots,c$
4. Calculate the topics : $\boldsymbol{\mu}_i = \max(0, decoder(\tilde{\boldsymbol{\mu}}_i)), i = 1,2,\ldots,c$

---

## 4. Results and Discussion

To examine the performance of DFCM, we apply the method to extract topics on two datasets, i.e., English email and Indonesian news. To prepare the textual data for the topic detection methods, we executed two main processes, i.e., cleaning and vectorizing. Firstly, we converted all words into lowercase, erase words containing domains such as www.* or https://*, and words containing @username, and erased # in #words. To standardize words with non-standard spelling, we replaced two or more repeating letters with only two occurrences. Stopwords and words with low-frequency terms occurring in fewer than *t* of the total *m* documents were also excluded, where the *t* threshold was set to *max*(10; *m*/1000). Finally, we use the term frequency-inverse document frequency for weighting.

Given a tweet collection, the topic detection methods produce topics represented by its top 10 most frequent words. The standard quantitative method to measure the interpretability of the topics is topic coherence. In our simulations, we use one of the topic coherence measures called TC-W2V (O'Callaghan, Greene, Carthy, & Cunningham, 2015). Suppose a topic $t$ consists of $n$ words that are $\{t_1, t_2, \ldots, t_n\}$, the TC-W2V of the topic $t$ is

$$TC - W2V(t) = \frac{1}{\binom{n}{2}} \sum_{j=2}^{n} \sum_{i=1}^{j-1} \text{similarity}(wv_j, wv_i) \qquad (11)$$

where $wv_j$ and $wv_i$ are vectors of word $t_j$ and $t_i$ constructed by a *word2vec* model.

The simulations are conducted in a Windows-based Python environment. For EFCM and DFCM parameters, we set the fuzzification constant $f = 1.1$, the maximum number of iteration $T = 1000$, and the threshold $\varepsilon = 0.005$. As mentioned before, the setting of the fuzzification constant is quite intuitive. The small fuzzification constant means that each tweet may contain a small number of topics. On the other hand, the bigger the fuzzification constant implies that each textual data may contain more topics. We initialize the centroids of FCM for both EFCM and DFCM using the best of the 10-run k-means clustering. In DFCM, DAE architecture consists of three symmetrical layers with each layer consisting of 500, 500, 2000 neurons. We implement this representation learning using Python-based Keras[1]. On the other hand, we use truncatedSVD implementation of scikit-learn for dimension reduction in EFCM (Pedregosa et al., 2011). Finally, we need to tune the rest parameters, i.e., the lower dimension $p$ and the number of topics $c$, for EFCM and DFCM.

This simulation also compares DFCM with two standard topic detection methods: latent Dirichlet allocation (LDA) and nonnegative matrix factorization (NMF). We use the LDA and NMF implementations provided by scikit-learn (Pedregosa et al., 2011). The LDA algorithm uses the batch variational Bayes method for training LDA. Two parameters are usually optimized for this training method, namely $\alpha$ and $\eta$. $\alpha$ control the mixture of topics for a specific document. A smaller $\alpha$ means the document will likely have less of a mixture of topics. $\eta$ control the distribution of words per topic. The larger $\eta$ means the topic will likely have more words. To optimize both parameters, we use a hyperparameter grid and run an algorithm for each combination [0.01, 0.1, 0.25, 0.5, 0.75, 1].

For NMF, the data vectors are normalized to unit length. The implementation of NMF uses a coordinate descent algorithm. There is no parameter we optimize for this algorithm. To reduce the instability of random initialization, the NNDSVD initialization is performed.

3.1 Enron – An English Email Dataset

The first dataset is Enron consisting of approximately 500,000 emails generated by employees of the Enron Corporation[2]. The Federal Energy Regulatory Commission obtained it during its investigation of Enron's collapse. To calculate the TC-W2V of the extracted topics, we use a pre-trained word2vec model trained on the Google News dataset for the English email dataset. The model contains 300-dimensional vectors for 3 million words and phrases[3].

---

[1] https://keras.io
[2] https://www.cs.cmu.edu/~./enron/
[3] https://code.google.com/archive/p/word2vec/

First, we analyze the effect of the number of DAE training epochs on the coherence scores of DFCM. Using a batch size of 256, the coherence scores for several epoch sizes are given in Figure 3. First, the number of epochs is set to 100. Then, this number of epochs is increased to 400 and 1000. The average coherence score of DFCM fluctuates and increases when the number of epochs is increased from 100 to 400. However, the average coherence score tends to decrease when the number of epochs is increased to 700. If we choose 400 as the number of epochs, then DFCM gives the mean coherence scores of 0.1771, 22% better than EFCM.

Figure 4 provides simulation results similar to Figure 3, but for 10-dimensional representation. Compared to the five-dimensional representation, the mean coherence scores of DFCM fluctuate only slightly as the number of epochs is increased from 100, 400, and 700. DFCM gives the mean coherence scores of 0.1896 when the number of epochs is 400. Like the 5-dimensional data representation, DFCM still provides a better mean coherence score than EFCM, which is about 5% better.

Figures 3 and Figure 4 show that the 10-dimensional representation of data is more suitable for both the DFCM and EFCM methods. DFCM with the ten-dimensional representation gives the mean coherence scores 7% better than DFCM with the five-dimensional representation. Meanwhile, EFCM with the ten-dimensional representation provides an average coherence score of 26% better than EFCM with the five-dimensional representation.

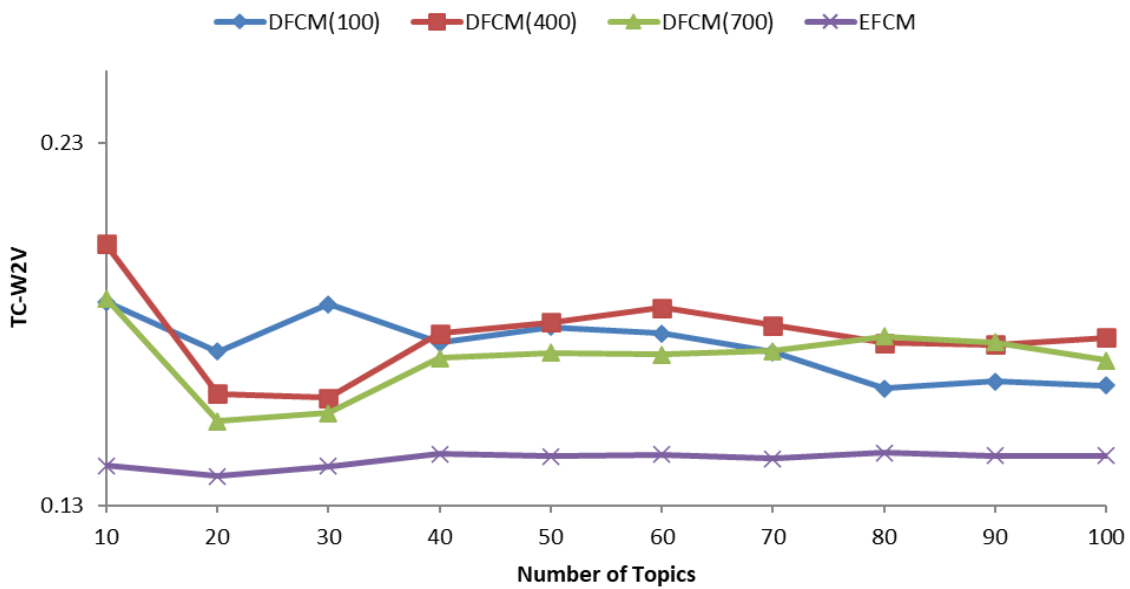

Figure 3. Coherence scores in terms of TC-W2V for the Enron dataset on the number of topics 10, 20, …, 100 when the lower-dimensional representation is set to five. DFCM(100), DFCM(400), DFCM(700) mean the number of epoch of deep autoencoders are set to 100, 400, 700, respectively

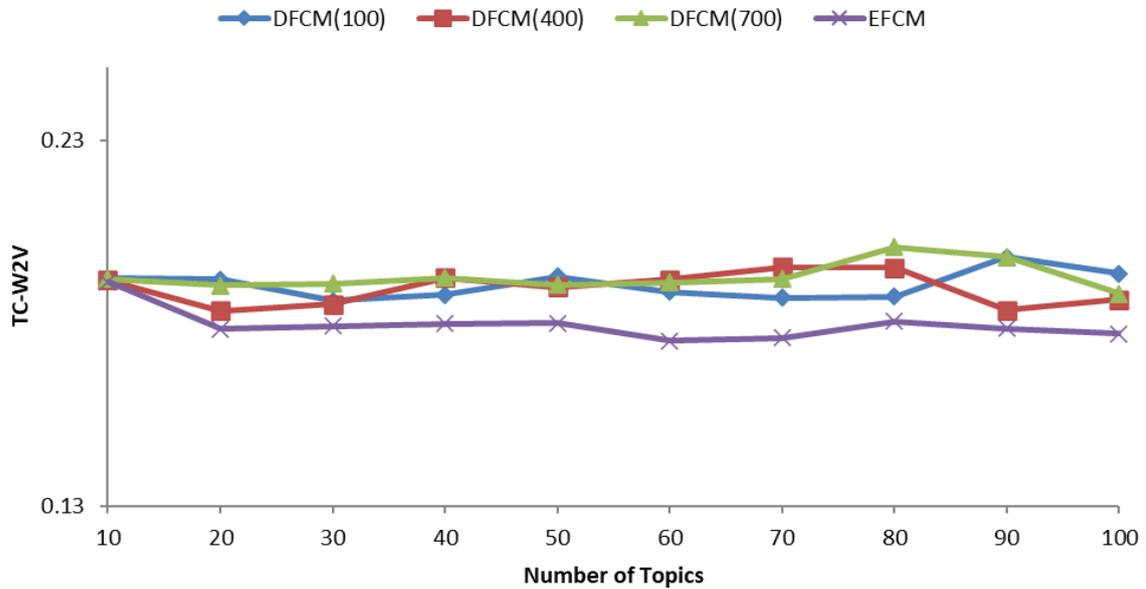

Figure 4. Coherence scores in terms of TC-W2V for the Enron dataset on the number of topics 10, 20, …, 100 when the lower-dimensional representation is set to ten. DFCM(100), DFCM(400), DFCM(700) mean the number of epoch of deep autoencoders are set to 100, 400, 700, respectively

Furthermore, we also provide a comparison between the DFCM method with two other standard methods, namely NMF and LDA. Figure 5 includes coherence scores of 4 topic detection methods for the number of topics 10, 20, ..., 100. First, Figure 5 confirms the previous simulation results that EFCM provides coherence scores between NMF and LDA. From Figure 5, we can also see that DFCM can reach a coherence score that is slightly better than NMF in almost all number of topics. Only on the topic number of 10, NMF gives a significantly better coherence score. For all number of topics, DFCM provides better mean coherence scores of 3%, 7%, 34% than NMF, EFCM, LDA, respectively.

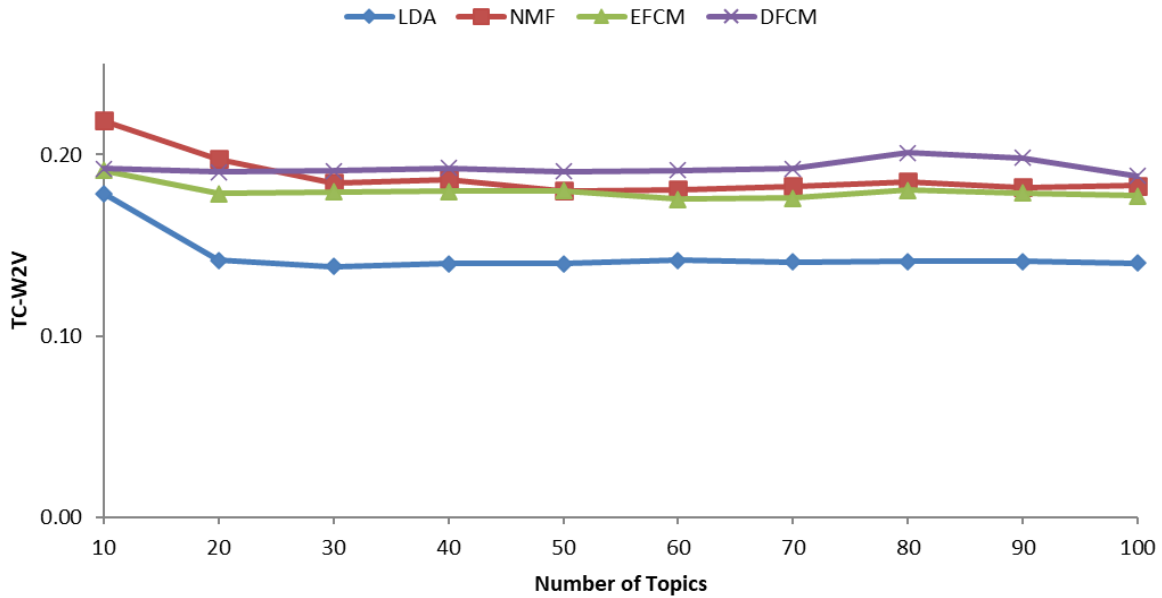

Figure 5. The comparison of coherence scores in terms of TC-W2V for LDA, NMF, EFCM, and DFCM on the number of topics 10, 20, …, 100 for the Enron dataset.

### 3.2 Berita – An Indonesian News Dataset

The second dataset is Berita, consisting of 50304 digital Indonesia news articles shared online through Twitter by nine Indonesian news portals widely known in Indonesia. These are Antara (antaranews.com), Detik (detik.com), Inilah (inilah.com), Kompas (kompas.com), Okezone (okezone.com), Republika (republika.co.id), Rakyat Merdeka (rmol.co), Tempo (tempo.co) and Viva (viva.co.id). The news articles contain published dates, titles, and some first sentences of contents. We construct the *word2vec* model using a corpus consisting of 750000 Indonesian documents from wiki, news, and tweets to measure the TC-W2V of the extracted topics. Unlike the *word2vec* model for the first English dataset, we train the Berita dataset to this *word2vec* model. Therefore, all vocabularies of the Berita dataset exist in the *word2vec* model.

    The simulations for the Berita dataset are given in Figure 6, Figure 7, and Figure 8. Figure 6 is a simulation to see the effect of the number of epochs in the DAE learning to coherence scores of DFCM for the five-dimensional representation. Meanwhile, Figure 7 is a simulation to see the effect of the number of epochs in the DAE learning to coherence scores of DFCM for the 10-dimensional representation. The initial number of epochs is 50 and is increased to 100 and then 400. In general, increasing the number of epochs makes the coherence score lower for most topics. For the epoch number of 400, DFCM even provides a coherence score below the EFCM in almost all number of topics. The same conditions are seen for the 10-dimensional representation in Figure 7.

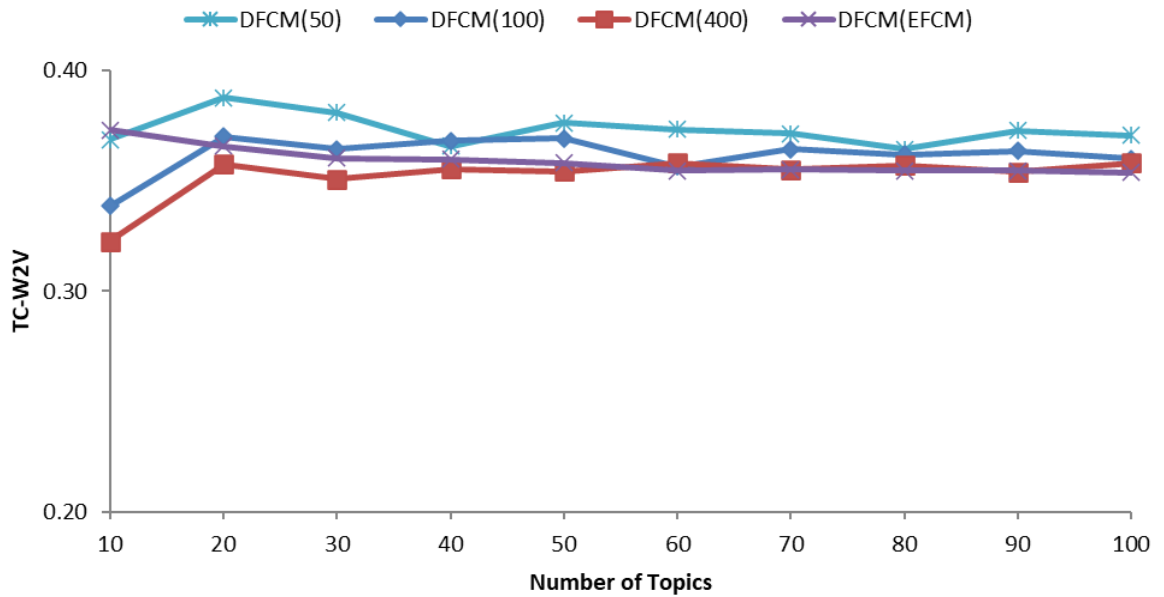

Figure 6. Coherence scores in terms of TC-W2V for the Berita dataset on the number of topics 10, 20, …, 100 when the lower-dimensional representation is set to five. DFCM(50), DFCM(100), DFCM(400) mean the number of epoch of deep autoencoders are set to 50, 100, 400, respectively

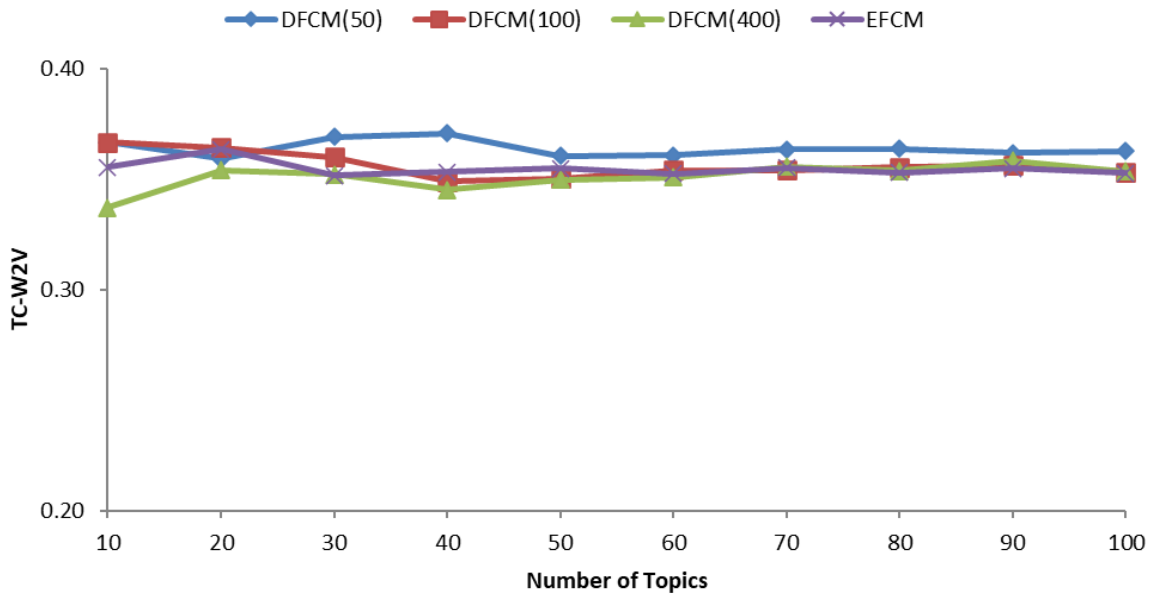

Figure 7. Coherence scores in terms of TC-W2V for the Berita dataset on the number of topics 10, 20, …, 100 when the lower-dimensional representation is set to ten. DFCM(50), DFCM(100), DFCM(400) mean the number of epoch of DAE are set to 50, 100, 400, respectively

If we use 50 epochs for DAE learning, then DFCM gives an average coherence score of 0.3730 for the five-dimensional representation. Meanwhile, EFCM provides an average coherence score of 0.3589. This means that DFCM achieves a slightly higher average coherence score than EFCM, which is about 4% better. A similar result is shown for the 10-dimensional representation where the DFCM gives an average coherence score of about 3%, slightly higher than the EFCM. From Figure 6 and Figure 7, we can also conclude that the five-dimensional representation provides a slightly better coherence score for both DFCM and EFCM.

Figure 8 provides a comparison of DFCM on the Berita dataset with two other standard topic detection methods, namely NMF and LDA. In this comparison, both DFCM and EFCM use the five-dimensional representation. Figure 8 shows that DFCM, EFCM, NMF, and LDA provide mean coherence scores of 0.3730, 0.3588, 0.3560, and 0.2815, respectively. These results indicate that DFCM can better achieve an average coherence score than EFCM, NMF, and LDA. However, NMF still gives a better coherence score for the smallest number of topics. 10. In this Berita dataset, NMF and EFCM provide almost the same means coherence score.

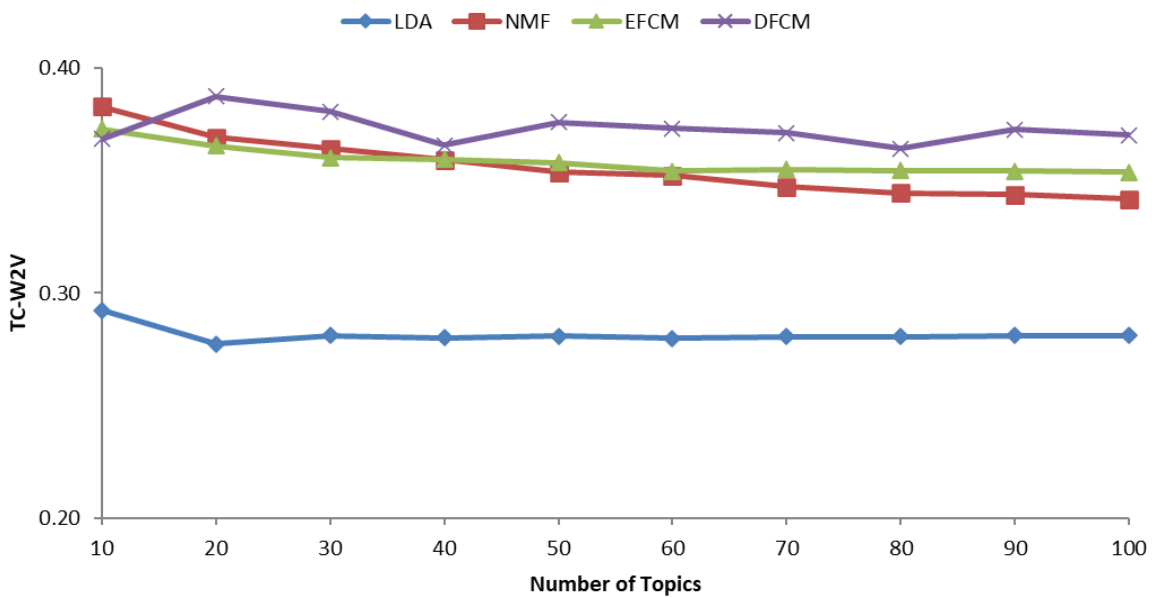

Figure 8. The comparison of coherence scores in terms of TC-W2V for LDA, NMF, EFCM, and DFCM on the number of topics 10, 20, …, 100 for the Berita dataset.

3.3 Discussion

In the previous sub-chapter, the simulations show that DFCM achieves better mean coherence scores than EFCM. For the Enron dataset, DFCM provides mean coherence scores that are 7% better than the EFCM and 4% more than the EFCM for the News dataset. The main difference between these two methods is the lower-dimensional representation learning process where DFCM uses DAE, while EFCM uses truncatedSVD. DAE and truncatedSVD produce different lower-dimensional representations. TruncatedSVD creates lower-dimensional representation with orthogonal dimensions or features. Meanwhile, DAE produces lower-dimensional representations with dimensions or features that are not orthogonal. Topics generally consist of words that are not necessarily orthogonal, especially in the meaning of the words. Also, DAE implements denoising processes implicitly to produce these lower-dimensional representations. Thus, each of these lower-dimensional characteristics will more or less affect the resulting mean coherence scores.

DFCM also provides a higher average coherence score compared to NMF. DFCM achieved a 3% better average coherence score for the Enron dataset and a 5% better average coherence score for the News dataset. In contrast to EFCM, DFCM and NMF provide lower-dimensional representation with non-orthogonal dimensions or features. NMF carried out a topic extraction process in the original space, which consisted of words. Thus, the resulting topics can be directly interpreted, and their coherence scores are calculated. Meanwhile, DFCM extracts the topics in the lower-dimensional space and must be transformed back to the original space so that the extracted topics can be interpreted, and coherence scores are calculated. However, DFCM makes it possible to process a better representation for textual data that generally has a lot of noise and variation. Thus, the success of DFCM in achieving better coherence scores is mainly because DFCM processes textual data with better representations. The same condition for LDA, where LDA performs the topic extraction process in the original space, consists of words.

Deep learning is currently a popular supervised learning approach, especially for unstructured data such as images and text. Deep learning integrates the feature extraction process with classification or regression processes. In the context of unsupervised learning, deep autoencoder is a popular deep learning method for representation learning. This method makes it possible to do the denoising process while reducing dimensions to produce better lower-dimensional representation. However, the integration of representation learning methods with an unsupervised learning problem such as topic detection is still an opportunity to continue to be developed.

## 5. Conclusions

DFCM is a topic detection method that combines deep autoencoder for representation learning and fuzzy c-means for topic extraction. Therefore, deep autoencoder-fuzzy c-means makes it possible to process a better representation for textual data that generally has a lot of noise and variation. Unlike EFCM, DFCM extracts topics from lower-dimensional representations with dimensions or features that are not orthogonal. This representation is more realistic to represent the topics. Our simulation shows that DFCM gives a higher accuracy in terms of the coherence score than EFCM and the two standard methods, i.e., NMF and LDA.

**Acknowledgement**
This paper was supported by Universitas Indonesia under PDUPT 2020 grant. Any opinions, findings, conclusion, and recommendations are the authors' and do not necessarily reflect those of the sponsor.